\documentclass[aps,prl,twocolumn,showpacs,showkeys,footinbib,amsmath,amssymb,superscriptaddress]{revtex4-2}

\usepackage{graphicx,color}
\usepackage{verbatim}
\usepackage{amssymb}   
\usepackage{amsmath}
\usepackage{amsfonts}
\usepackage{mathdots}
\usepackage{braket}
\usepackage{bm}
\usepackage{float}
\usepackage{bbm}

\usepackage{soul}
\usepackage{pifont}

\PassOptionsToPackage{colorlinks=true,
	linkcolor=blue,
	citecolor=blue,
	urlcolor=blue}{hyperref}
\usepackage{hyperref}

\begin{document}
          
\title{Towards Scalable Braiding: Topological Superconductivity Unlocked under Arbitrary Magnetic Field Directions in Curved Planar Josephson Junctions}

\author{Richang Huang}
\affiliation{Eastern Institute for Advanced Study, Eastern Institute of Technology, Ningbo, Zhejiang 315200, China}

\author{Yongliang Hu}
\affiliation{Eastern Institute for Advanced Study, Eastern Institute of Technology, Ningbo, Zhejiang 315200, China}
\affiliation{School of Physics and Astronomy, Shanghai Jiao Tong University, Shanghai 200240, China}

\author{Xianzhang Chen}
\affiliation{Eastern Institute for Advanced Study, Eastern Institute of Technology, Ningbo, Zhejiang 315200, China}

\author{Peng Yu}
\affiliation{University of Chinese Academy of Sciences, Beijing 100049, China}

\author{Siwei Tan}
\affiliation{College of Computer Science, Zhejiang University, Hangzhou 310058, China}

\author{Igor \v{Z}uti\'c}
\affiliation{Department of Physics, University at Buffalo, State University of New York, Buffalo, New York 14260, USA}

\author{Tong Zhou}
\email{tzhou@eitech.edu.cn}
\affiliation{Eastern Institute for Advanced Study, Eastern Institute of Technology, Ningbo, Zhejiang 315200, China}

\date{\today}

\begin{abstract}

The non-Abelian statistics of Majorana zero modes (MZMs) are central to fault-tolerant topological quantum computing. Planar Josephson junctions provide a particularly versatile platform for realizing robust topological superconductivity hosting MZMs over a broad parameter space. However, it is generally believed that such topological superconductivity is restricted to a narrow range of in-plane magnetic field orientations, posing a major obstacle to scalable and noncollinear junction-network architectures. Here, we uncover that the apparent suppression of MZMs under misaligned fields does not arise from the destruction of topological superconductivity itself, but instead originates from emergent shifted bulk states at other momenta that obscure the global excitation gap and MZMs. By introducing spatial modulations along the junction to scatter and gap out these bulk states, we restore a global topological gap and recover MZMs for arbitrary in-plane magnetic field orientations. Remarkably, such modulations can be naturally realized by transforming a straight junction into a curved geometry, rendering the topological gap robust against field misalignment and enabling MZMs survival in complex junction networks. Building on this robustness, we propose a scalable protocol for MZMs braiding and fusion using gate or superconducting-phase control, opening new routes toward scalable topological quantum computing.

\end{abstract}

\maketitle

The quest to identify systems capable of hosting and manipulating Majorana zero modes (MZMs) for fault-tolerant quantum computing~\cite{nayak2008non,sarma2015majorana,Alicea2011:NP,aasen2016milestones,laubscher2021majorana,flensberg2021engineered} has stimulated extensive exploration of condensed-matter platforms, among which heterostructures have emerged as a central focus~\cite{fu2008superconducting,qi2010chiral,Nadj-Perge2014:S,wang2018high,yan2018majorana,fu2021chiral,pan2024majorana,hu2024chiral,zhu2025alter,vzutic2019proximitized,Amundsen2024:RMP,qi2011topological}. Early theoretical proposals and many experimental efforts concentrated on one-dimensional (1D) semiconductor--superconductor systems~\cite{lutchyn2010majorana,oreg2010helical,alicea2010majorana,klinovaja2012transition,mourik2012signatures,rokhinson2012fractional,microsoft2025interferometric}. However, zero-bias conductance peaks~\cite{sengupta2001midgap,liu2012zero,pan2020physical} observed in such 1D systems may originate from mechanisms unrelated to MZMs~\cite{Chen2019:PRL,pan2020generic,DasSarma2021:PRB}, and the reliance on finely tuned parameters has hindered these experiments from providing unambiguous evidence~\cite{pan2020generic,DasSarma2021:PRB,Yu2021:NP}. Moreover, strictly 1D platforms face intrinsic limitations for MZMs braiding.

\begin{figure}[ht!]
\label{fig1}
\includegraphics[width=0.48\textwidth]{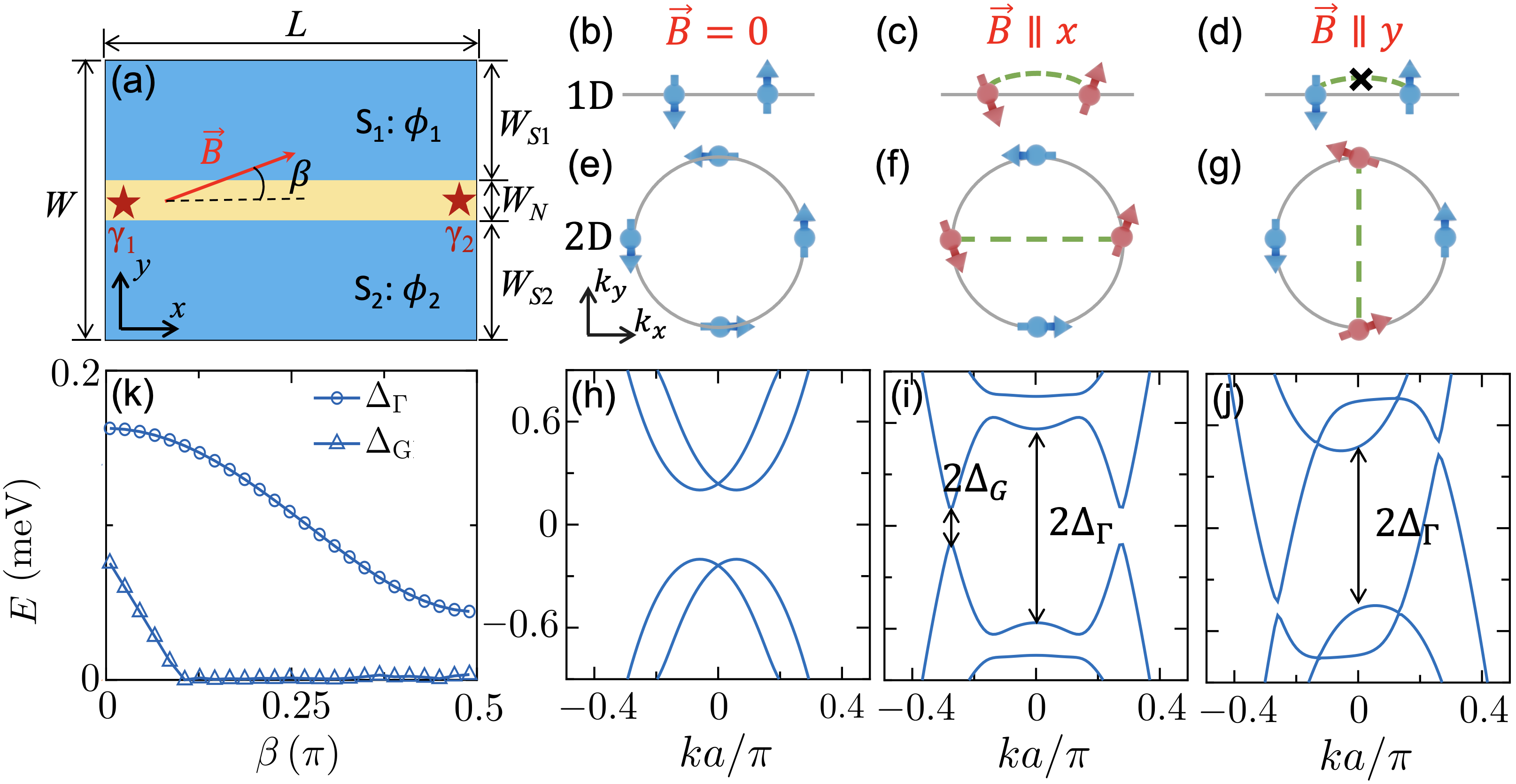}
\caption{(a) Schematic of the PJJ with $\vec{B}$ misaligned by $\beta$ relative to the interface between S (blue) and N (yellow) regions with phase difference, $\phi=\phi_1-\phi_2$, where MZMs (stars) can emerge at the ends of the N region. (b)–(d) Spin texture on the Fermi surface for a 1D system at $\vec{B}=0$, $\vec{B}\|x$, and $\vec{B}\|y$. (e)–(g) Same as (b)–(d), but for a PJJ (2D system). The canted spin texture (red) enabling effective $p-$wave pairing are indicated by the green dashed lines. (h)–(j) Corresponding PJJ spectra for (e)–(g), where $\Delta_{\Gamma}$ denotes the gap at $k=0$ and $\Delta_{G}$ the global gap. (k) $\beta$ dependence of $\Delta_{\Gamma}$ and $\Delta_{G}$, where the parameters are $L=6500$ nm, $W_{N}=100$ nm, $W_{S1}=W_{S2}=300$ nm, $\mu=2$ meV, $B=0.3$ T and $\phi=\pi$. Other parameters are specified in the main text.}
\end{figure}

Alternatively, planar Josephson junctions (PJJs), as depicted in Fig.~\hyperref[fig1]{1(a)}, constitute a particularly promising two-dimensional (2D) platform for realizing topological superconductivity (TSC)~\cite{hell2017two,pientka2017topological,stern2019fractional,scharf2019tuning,fornieri2019evidence,ren2019topological,Banerjee2023:PRB,dartiailh2021phase,pekerten2022anisotropic,zhou2020phase,zhou2022fusion,pekerten2024beyond,schiela2024progress}, which has been experimentally demonstrated to persist over a broad range of parameters~\cite{fornieri2019evidence,ren2019topological,dartiailh2021phase,Banerjee2023:PRB}. PJJs offer the flexibility to tune the superconducting phase bias~\cite{zhou2020phase} and gate voltages~\cite{zhou2022fusion}, enabling MZM control and demonstration of their non-Abelian statistics. Despite these advantages, the straight PJJ (SJJ) faces a significant limitation: TSC is highly sensitive to the orientation of the in-plane magnetic field, $\vec{B}$~\cite{dartiailh2021phase,pekerten2022anisotropic}, characterized by the misalignment angle $\beta$ relative to the interface between the superconducting (S) and normal (N) regions. Experiments have shown that even a small misalignment suppresses TSC and disrupts MZMs formation~\cite{dartiailh2021phase}. This pronounced sensitivity poses a practical challenge for scalable networks, where the $\vec{B}$ orientation can vary across different junctions~\cite{zhou2020phase,schiela2024progress}.

Motivated by this stringent constraint, we propose a universal and experimentally feasible strategy based on spatial modulation to unlock TSC in PJJs for arbitrary $\vec{B}$ orientations. We show that the frequently reported suppression of MZMs at finite $\beta$ does not imply that TSC itself is destroyed; rather, broken reflection symmetry tilts the spectrum and brings additional bulk modes from finite momenta into the low-energy window, obscuring both the global gap and the MZMs. We therefore propose to eliminate these masking states by transforming a SJJ into a curved PJJ (CJJ) using geometric lithography~\cite{zhou2020phase,zhou2022fusion,yu2025gate}, which selectively scatters and gaps out the unwanted modes while leaving the zone-center topology intact. This restores a full topological gap and robust MZMs, detectable through standard PJJ signatures such as critical-current minima and accompanying phase jumps~\cite{pientka2017topological,dartiailh2021phase}. By eliminating the bottleneck of precise $\vec{B}$ alignment, the resulting spatially engineered junction provides a practical route to gate- or phase-controlled braiding and fusion protocols in scalable, noncollinear networks.

To clarify the origin of the robustness against $\vec{B}$ misalignment, we contrast PJJs with conventional 1D nanowires. In 1D proposals~\cite{lutchyn2010majorana,oreg2010helical}, a perpendicular $\vec{B}$ component to the spin–orbit-coupling (SOC) is required to generate a canted spin texture that supports effective $p$-wave pairing [Figs.~\hyperref[fig1]{1(b)–1(d)}], fostering the perception that precise $\vec{B}$ alignment is essential for stabilizing MZMs~\cite{osca2014effects,rex2014tilting}. By analogy, PJJs are often assumed to require an in-plane $\vec{B}$ along the junction ($\beta=0$), with small transverse components ($\beta \neq 0$) rapidly suppressing MZMs, consistent with experiments~\cite{fornieri2019evidence,ren2019topological,dartiailh2021phase,Banerjee2023:PRB}. However, PJJs are intrinsically 2D: Rashba SOC winds in momentum space with both $x$ and $y$ components [Fig.~\hyperref[fig1]{1(e)}], so the SOC–B interplay yields a canted spin texture-and thus supports TSC~\cite{qi2011topological,Amundsen2024:RMP}—for essentially any in-plane $\vec{B}$ direction [Figs.~\hyperref[fig1]{1(f)} and~\hyperref[fig1]{1(g)}]. The key difference is therefore not the local topological criterion, but the global spectrum. At $\beta=0$, reflection symmetry protects a finite global gap and stabilizes MZMs; at $\beta=\pi/2$, this symmetry is broken, the bands tilt, and gap closings appear at finite momenta. Consequently, although the $k=0$ band structure (and thus the $\mathbb{Z}_2$ invariant $Q$) is unchanged [Fig.~\hyperref[fig1]{1(i)}], the global topological gap collapses and MZMs lose protection [Fig.~\hyperref[fig1]{1(j)}]. This is evident in Fig.~\hyperref[fig1]{1(k)}: $\Delta_{\Gamma}$ remains finite and topologically nontrivial for all $\beta$, whereas $\Delta_{G}$ is rapidly suppressed by shifted finite-momentum states. Because these states arise away from $k=0$, spatial modulation can selectively gap them out without affecting the underlying topology~\cite{tewari2012topological,qi2011topological}, thereby restoring the global gap and yielding MZMs that remain robust against $\vec{B}$ misalignment.

\begin{figure}
\includegraphics[width=0.48\textwidth]{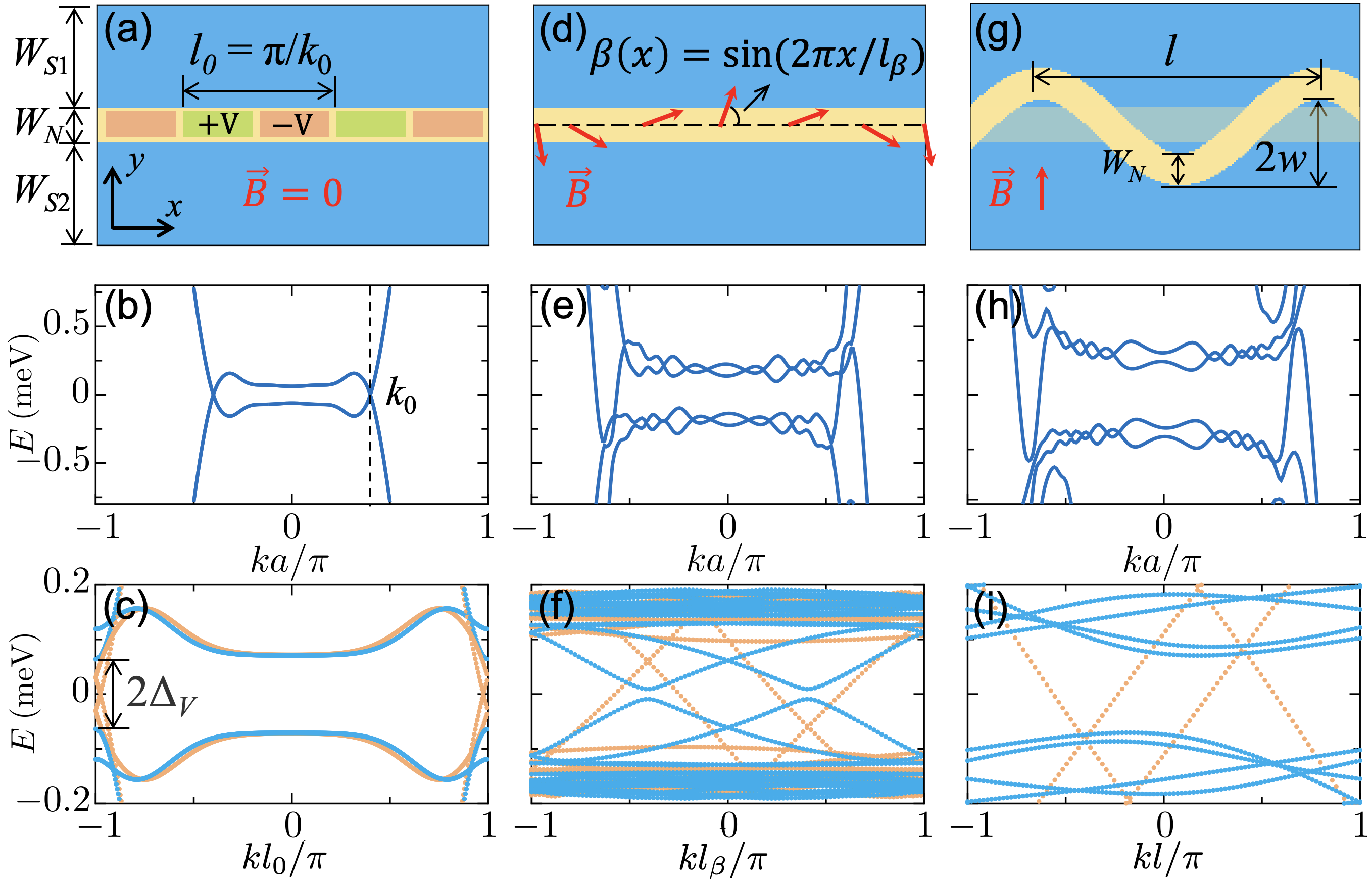}
\caption{\label{fig2} (a) Same as Fig.~\hyperref[fig1]{1(a)}, but with a gate array (orange and green) producing a periodic step modulation of the chemical potential with amplitude $V$ and period $l_0$. (b) Energy spectrum of (a) at $\mu=2.2$ meV, $V=0$ and $\vec{B}=0$, exhibiting a gap closing at momentum $k_0=0.36\pi/a$. (c) Folded spectrum of (b) (orange) in the mini Brillouin zone with superlattice period $l_0=\pi/k_0$; a finite $V$ opens a gap at $k_0$ (blue) (d) Same geometry as (a), but with a spatially noncollinear magnetic texture generated by a sinusoidal modulation of $\beta$ along the $x$ direction. (e) Energy spectrum of (d) for $\mu=7$ meV, $\vec{B}\|y$ ($B=0.4\,$T, $\beta=\pi/2$). (f) Same as (e), but with $\beta=\sin(2\pi x/l_\beta)$ ($l_\beta = 1300$ nm), which opens a gap (blue) compared to the uniform $\beta=\pi/2$ case (orange). (g) Same as (a), but for a CJJ with a sinusoidal N region of period $l=1300$ nm and amplitude $w=200$ nm. (h) Same as (e), but at $\mu=8$ meV. (i) Same as (h), but for the CJJ, where a global gap opens (blue) compared to the SJJ spectrum (orange). Other parameters are taken from Fig.~\hyperref[fig1]{1}.}
\end{figure}

To further demonstrate our proposal, we simulate spatially modulated PJJs using the Bogoliubov–de Gennes (BdG) Hamiltonian,

\begin{equation}
\label{ham}
\begin{aligned}
H= & {\left[\frac{p_x^2+p_y^2}{2 m^*}-\mu+V(x, y)+\frac{\alpha}{\hbar}\left(p_y \sigma_x-p_x \sigma_y\right)\right] \tau_z } \\
& -\frac{g^* \mu_B}{2} \vec{B} \cdot \vec{\sigma}+\Delta(x, y) \tau_{+}+\Delta^*(x, y) \tau_{-},
\end{aligned}
\end{equation}
where $p_{x,y}$ is the momentum, $m^*$ the effective mass, $\mu$ the chemical potential, and $\alpha$ the Rashba SOC strength. We denote by $\tau_i$ and $\sigma_i$ the Pauli and Nambu matrices acting in spin and particle–hole space, respectively, with $\tau_{\pm}=(\tau_x\pm i\tau_y)/2$. The proximity-induced pairing potential $\Delta(x,y)=\Delta_0 e^{i\phi_{i}}$ is present in the 2D electron gas beneath the S regions, where  $\Delta_0$ is the induced gap and $\phi_i$ the corresponding superconducting phase. $V(x,y)$ describes the local changes of $\mu$ due to the application of the mini-gate voltages. Throughout, we use parameters representative of well-established semiconductor–superconductor heterostructures~\cite{yan2023supercurrent,schiela2024progress}: $m^*=0.02m_0$ (with $m_0$ the electron mass), $g^*=26$, $\Delta_0=1$ meV, and $\alpha=20$ meV nm. The calculations are performed by numerically solving Eq.~\hyperref[ham]{(1)} on a discretized lattice using KWANT~\cite{Groth2014kwant}; further computational details are provided in the Supplemental Material (SM)~\cite{SM}.

\begin{figure*}
\includegraphics[width=17.2cm]{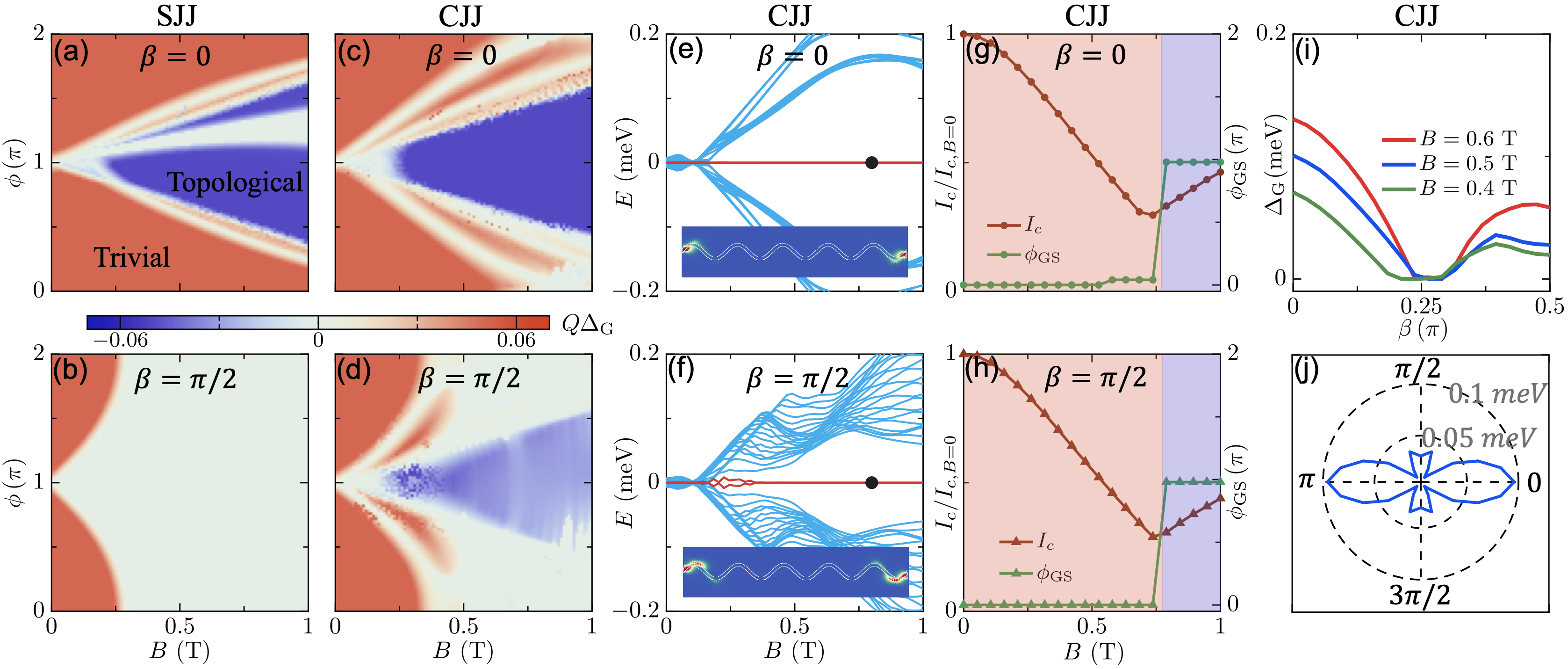}
\caption{\label{fig3} (a)-(b) Phase diagram of a SJJ, plotted as the product $Q\Delta_G$, for $\beta=0$ and $\beta=\pi/2$. (c)-(d) Same as (a)-(b) but for a CJJ, where the gapped topological region suppressed at $\beta=\pi/2$ in (b) is recovered in (d). (e)-(f) Energy spectra of the CJJ at $\phi=\pi$ for $\beta=0$ and $\beta=\pi/2$; insets show the probability density of the zero-energy modes at $B=0.8$ T (black dots). (g)-(h) Critical current $I_c$ (brown) and ground-state phase $\phi_{GS}$ (green) versus $B$ for the CJJ at $\beta=0$ and $\beta=\pi/2$. (i) $\beta$ dependence of $\Delta_G$ in the CJJ. (j) Polar map of $\Delta_G(\beta)$ at $B=0.5$ T. Other parameters are taken from Fig.~\hyperref[fig2]{2}.}
\end{figure*}

To analytically understand why the topological phase in a PJJ is insensitive to the $\vec{B}$ direction, it is instructive to first consider the SOC-free limit, where the spin in Eq.~\hyperref[ham]{(1)} depends only on $\vec{B}$ value and is therefore invariant under rotations of $\vec{B}$. In realistic devices SOC is present, but it vanishes at $k$=0. Because the topological transition is controlled by the band topology at $k$=0, one expects the topological condition to be independent of the field orientation. We confirm this expectation by deriving the phase boundary from Eq.~\hyperref[ham]{(1)} using scattering theory~\cite{beenakker1991universal} (see SM for details~\cite{SM}) as 

\begin{equation}\label{cd}
 \frac{g^*\mu_{B}B\, W_{N}}{2v_F} \pm \frac{\phi}{2}=\frac{\pi}{2}+\pi n,
\end{equation}
where $v_F$ is the Fermi velocity. This condition, confirmed by our numerical calculations [Fig.~\hyperref[fig3]{3(a)}], is explicitly independent of the $\vec{B}$ directions: it not only reproduces earlier results in the $\beta$ = 0 limit~\cite{pientka2017topological} but also remains valid for arbitrary $\beta$, including $\beta$=$\pi$/2.

For a finite $\beta$, however, the TSC can be effectively hidden because shifted bulk modes at finite momenta collapse the global gap $\Delta_{G}$, even though the zone-center gap $\Delta_{\Gamma}$ remains finite, as shown in Fig.~\hyperref[fig3]{3(b)}. Restoring a global gap therefore requires selectively gapping out these unwanted finite-$k$ states. For a spinless gapless mode at $k=k_0$ Fig.~\hyperref[fig2]{2(b)}], a periodic step potential provides an efficient route via Bragg scattering~\cite{ashcroft1976solid}: the minimal period $l_0=\pi/k_0$ backscatters the mode and opens a gap, as confirmed numerically for a gate-array-modulated SJJ [Figs.~\hyperref[fig2]{2(a)-(c)}]. For spinful gapless modes, an electrostatic modulation alone is generally insufficient, since it does not mix spins. Instead, spatially varying magnetic textures—well known to induce or enhance TSC in a broad range of platforms ~\cite{choy2011majorana,Klinovaja2013:PRL,Nadj-Perge2014:S,fatin2016wireless,zhou2019tunable,
hess2022prevalence,desjardins2019synthetic,Marra2017:PRB,Yang2016:PRB,Gungordu2018:PRB,Palacio-Morales2019:SA,mohanta2021skyrmion,gungordu2022majorana}—can provide the required spin-dependent scattering. Indeed, for a spinful gapless spectrum in the SJJ [Fig.~\hyperref[fig2]{2(e)}], a periodic local misalignment $\beta(x)$ [Fig.~\hyperref[fig2]{2(d)}] reopens the gap [Fig.~\hyperref[fig2]{2(f)}]. In realistic PJJs under misaligned fields, both spinless and spinful finite-k crossings may occur, motivating a single, experimentally practical design that incorporates both mechanisms. This is naturally achieved by geometric modulation: for example, in a sinusoidally modulated curved junction [Fig.~\hyperref[fig2]{2(g)}], the curved normal channel experiences an effective periodic potential modulation while the globally applied $\vec{B}$ converted into a spatially varying local field texture. Our calculations show that the shifted gapless states present in the SJJ [Fig.~\hyperref[fig2]{2(h)}] are efficiently gapped out in the CJJ [Fig.~\hyperref[fig2]{2(i)}], thereby restoring a global gap and rendering the previously masked MZMs visible and topologically protected as shown in Fig.~\hyperref[fig3]{3}. Remarkably, the same CJJ modulation also enhances TSC under aligned magnetic fields, as shown in Fig.~\hyperref[fig3]{3(c)}, consistent with previous studies~\cite{laeven2020enhanced,paudel2021enhanced,woods2020enhanced,adagideli2014effects,melo2023greedy,schirmer2024topological}. Together, these results demonstrate that experimentally accessible geometric engineering provides a powerful and general route to unlock and stabilize topological superconductivity in PJJs for arbitrary magnetic-field orientations.

\begin{figure*}
\includegraphics[width=17.2cm]{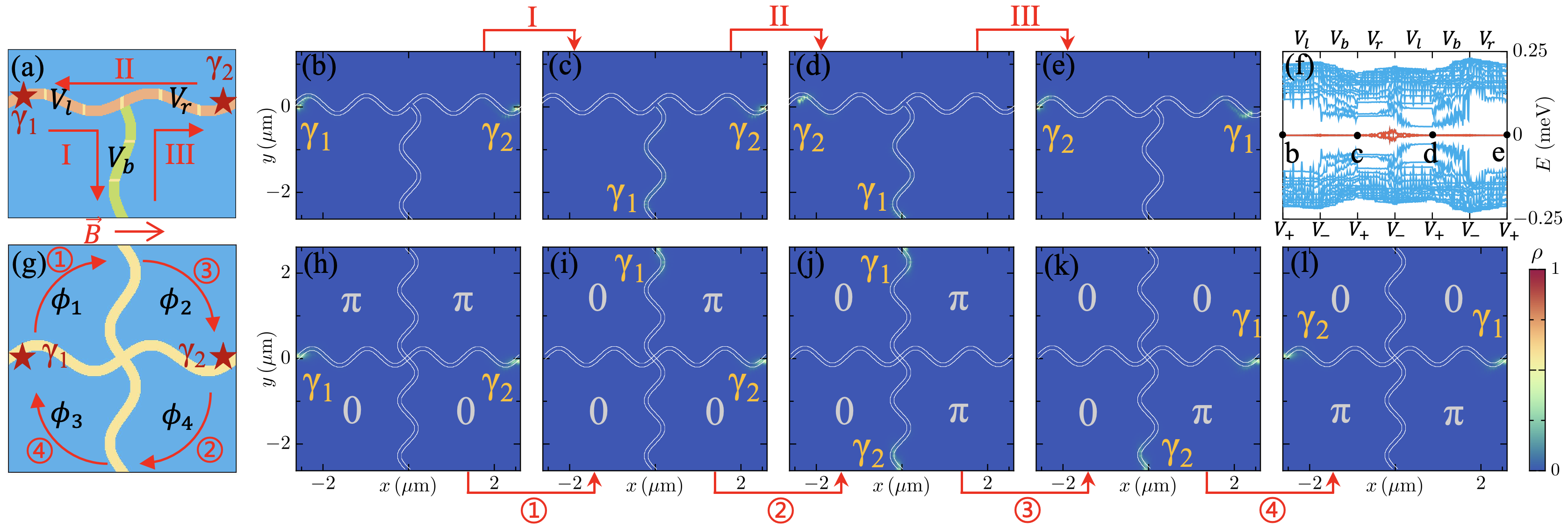}
\caption{\label{fig4} (a) Schematic of gate-controlled MZMs braiding in a CJJ T junction, where the superconducting phase difference between adjacent S regions is fixed at $2\pi/3$. The chemical potential in each branch is tuned by mini-gate voltages ($V_l$, $V_r$, and $V_b$) between topological ($V_+$, orange) and trivial ($V_-$, green) regimes, enabling MZMs exchange through the three steps I–III indicated by red arrows. (b)–(f) Evolution of the calculated MZMs probability density $\rho$ (color scale) and the corresponding energy spectrum during steps I–III, with $V_+=4$ meV and $V_-=10$ meV. (g) Schematic of phase-controlled MZMs exchange in a CJJ cross junction, implemented by tuning the superconducting phases $\phi_{1-4}$ according to steps \ding{192}-\ding{195}. (h)–(l) Evolution of the calculated $\rho$ during steps \ding{192}–\ding{195}. Other parameters are taken from Fig.~\hyperref[fig3]{3}}
\end{figure*}

Having established that TSC and MZMs persist in the CJJs even for $\beta = \pi/2$, it is instructive to compare this case with the conventional aligned-field configuration $\beta = 0$. Once the topologically irrelevant finite-momentum states are gapped out by geometric modulation, the resulting topological phase diagrams, energy spectra, and MZM wave-function distributions for $\beta = 0$ and $\beta = \pi/2$ become nearly identical, as shown in Figs.~\hyperref[fig3]{3(c)-(f)}. This close correspondence reflects the fundamental insensitivity of the underlying topological properties in PJJs to the $\vec{B}$ orientation. Since robust TSC is recovered at both limiting angles, $\beta = 0$ and $\beta = \pi/2$, it is natural to expect stability at intermediate orientations. This expectation is confirmed by the $\beta$-dependent  $\Delta_{G}$ shown in Figs.~\hyperref[fig3]{3(i)} and \hyperref[fig3]{3(j)}, which remains finite for essentially arbitrary $\beta$, underscoring the suitability of CJJs for scalable braiding in noncollinear junction networks.

Beyond establishing the existence of MZMs, their experimental identification remains a central challenge. In SJJ, a pronounced minimum in the critical current accompanied by an abrupt $\pi$ jump in the ground-state phase difference serves as a distinctive signature of TSC, beyond conventional zero-bias conductance peaks, and has been observed experimentally~\cite{dartiailh2021phase}. We find that these hallmark features persist in the CJJ for both $\beta=0$ and $\beta=\pi/2$ [Figs.~\hyperref[fig3]{3(g)} and \hyperref[fig3]{3(h)}], providing experimentally accessible and unambiguous probes of the restored topological superconducting phase.

The non-Abelian statistics of MZMs constitute a smoking-gun signature of their existence and form the foundation of topological quantum computation~\cite{nayak2008non,Alicea2011:NP,sarma2015majorana,aasen2016milestones,laubscher2021majorana,flensberg2021engineered,bedow2024simulating}. However, the key operations that reveal this property—braiding and fusion—have not yet been experimentally realized, representing a central challenge in the field~\cite{schiela2024progress}. In conventional semiconductor nanowires, a standard route to braiding is to use mini gates in a T junction to move and exchange MZMs~\cite{Alicea2011:NP,aasen2016milestones}. Such approaches, however, overlook the fact that $\vec{B}$ is applied globally across the device, inevitably producing locally misaligned fields in junction networks, which can suppress topological superconductivity and destroy MZMs~\cite{osca2014effects,rex2014tilting}. This limitation poses a fundamental obstacle to scalable architectures. In contrast, our proposed CJJs here naturally overcome this constraint, as its TSC is robust against magnetic-field misalignment.

To demonstrate this advantage and realize MZMs braiding without constraints on $\vec{B}$ orientation, we propose a gate-controlled braiding protocol based on a CJJ T junction, shown in Fig.~\hyperref[fig4]{4}. The local chemical potentials of the three junction arms are tuned by gate voltages $(V_l,V_r,V_b)$, implemented via mini gates~\cite{zhou2022fusion}, to switch each arm between topological ($V_+$) and trivial ($V_-$) regimes. As illustrated in Figs. ~\hyperref[fig4]{4(b)} and ~\hyperref[fig4]{4(c)}, the MZM $\gamma_1$ is adiabatically transferred from the left arm to the bottom arm by tuning $(V_l,V_r,V_b)$ from $(V_+,V_+,V_-)$ to $(V_-,V_+,V_+)$. Subsequently, $\gamma_2$ is moved from the right arm to the left arm [Fig.~\hyperref[fig4]{4(d)}], followed by the transfer of $\gamma_1$ from the bottom arm to the right arm [Fig.~\hyperref[fig4]{4(e)}]. Together, these steps implement a complete braiding operation, carried out adiabatically within the restored global gap $\Delta_G$, as confirmed by the evolution of the energy spectrum shown in Fig.~\hyperref[fig4]{4(f)}.

PJJs also provide additional, experimentally accessible control knobs, most notably the superconducting phase bias~\cite{zhou2020phase}. Leveraging this flexibility, we design a CJJ cross junction [Fig.~\hyperref[fig4]{4(g)}] and implement MZMs braiding purely via phase control. The exchange is carried out through a sequence of phase-tuning steps, as shown in Figs.~\hyperref[fig4]{4(h)-(l)}. Because the CJJ geometry renders TSC robust to field orientation, $\vec{B}$ misalignment is no longer a bottleneck for scalable braiding protocols~\cite{zhou2020phase}. In addition, the cross-junction naturally enables MZMs fusion and double-braiding operations, essential primitives for topological quantum computation~\cite{matos2017tunable,liu2019flux,martin2020double}, with the non-Abelian nature directly revealed by the dependence on operation order (see SM~\cite{SM}).

While our work is motivated by semiconductor PJJs, where multiple topological signatures have been experimentally demonstrated~\cite{fornieri2019evidence,ren2019topological,dartiailh2021phase,Banerjee2023:PRB} and uncovered 2DEG regions enable effective gate and phase control~\cite{zhou2020phase,zhou2022fusion}, the principle of using geometric curvature to gap out unwanted states and enhance topological superconductivity is far more general. This strategy can be straightforwardly extended to other Josephson-junction platforms~\cite{li2021topological,zhang2025exchange,wei2016induced}, including those based on topological insulators and semimetals, where finite-momentum modes can similarly limit the global gap. It is also relevant to systems with nonstandard SOC, such as Ge-based PJJs~\cite{Luethi2023:PRB,hinderling2024direct,leblanc2025gate} with dominant cubic-in-momentum SOC~\cite{Amundsen2024:RMP}, where Majorana modes may acquire higher-angular-momentum character~\cite{Alidoust2021:PRB}. More broadly, the concepts developed here apply to generic 2DEG platforms beyond PJJs including electrostatically defined multiband nanowire networks that have recently shown rapid progress toward topological quantum computation~\cite{microsoft2025interferometric}. In these platforms, the SOC differs qualitatively from that in earlier epitaxially defined 1D nanowires~\cite{mourik2012signatures}, opening additional opportunities for geometric optimization along the lines proposed here. Finally, we expect our proposed geometric modulation to be equally valuable for enhancing the robustness of topological phases in nonsuperconducting systems~\cite{zhou2016quantum,Zhou2021:PRL,qi2011topological}.

\begin{acknowledgements} This work is supported by the National Natural Science Foundation of China (12474155), the Zhejiang Provincial Natural Science Foundation of China (LR25A040001), and U.S. ONR under award MURI N000142212764 (I.\v{Z}.). The computational resources for this research were provided by the High Performance Computing Platform at the Eastern Institute of Technology, Ningbo. 
\end{acknowledgements}

\bibliography{main_bibliography}

\end{document}